\documentclass[final,5p,twocolumn]{elsarticle}
\usepackage{graphicx}
\usepackage{bm}
\usepackage{amsmath}
\usepackage{amssymb}
\usepackage{appendix}
\usepackage{subfigure}
\usepackage{xcolor, soul, color}
\sethlcolor{red}

\begin{document}


\title{Current dependence of the low bias resistance of small capacitance Josephson junctions} 

\author{Venkat Chandrasekhar}
\affiliation{Department of Physics, Northwestern University, Evanston, Illinois. 60208, USA}%

\date{\today}

\begin{abstract}
The dc current-voltage characteristics of small Josephson junctions reveal features that are not observed in larger junctions, in particular, a switch to the finite voltage state at current values much less than the expected critical current of the junction and a finite resistance in the nominally superconducting regime.  Both phenomena are due to the increased sensitivity to noise associated with the small capacitance of the Josephson junction and have been extensively studied a few decades ago.  Here I focus on the current bias dependence of the differential resistance of the junction at low current bias in the nominally superconducting regime, using a quantum Langevin equation approach that enables a physically transparent incorporation of the noise environment of the junction.  A similar approach might be useful in modeling the sensitivity of superconducting qubits to noise in the microwave regime.  
\end{abstract}

\maketitle
Josephson junctions (JJs) form the heart of superconducting qubits, arguably one of the leading platforms for quantum computing and quantum sensing applications.  Indeed, JJs and devices incorporating JJs were one of the earliest systems used to study macroscopic quantum tunneling and macroscopic quantum coherence \cite{voss_macroscopic_1981, devoret_measurements_1985, clarke_quantum_1988,ambegaokar_quantum_1982, fisher_quantum_1985, clarke_macroscopic_1986, blackburn_survey_2016}.  In particular, the role of noise in the environment and its effect on quantum tunneling were extensively studied almost four decades ago, where it was found that the dissipation introduced by the noise reduced the probability of tunneling \cite{ambegaokar_quantum_1982, caldeira_quantum_1983}.  This analysis was applied to studies of the switching current of JJs as a function of temperature \cite{voss_macroscopic_1981, devoret_measurements_1985, clarke_quantum_1988}, and later to understand the appearance of a finite resistance in the nominally zero-voltage regime in very small junctions \cite{iansiti_charging_1987, iansiti_crossover_1988, kautz_noise-affected_1990, martinis_classical_1989, joyez_josephson_1999, ivanchenko_josephson_1969, biswas_effect_1970, hu_low-voltage_1992},  The latter phenomena has been modeled using Monte Carlo methods primarily in the regime where thermal activation is dominant \cite{kautz_noise-affected_1990, buttiker_thermal_1983, joyez_josephson_1999}.  Here I focus on the low temperature regime where quantum tunneling is dominant, using the Wentzel-Kramer-Brillouin (WKB) approximation \cite{merzbacher} to calculate the tunneling rate.  This approach gives better physical insight into the different factors affecting the zero-bias resistance, although the results still need to be calculated numerically.

\begin{figure}
    \centering
    \includegraphics[width=0.6\columnwidth]{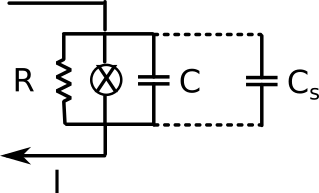}
    \caption{Equivalent circuit of a Josephson junction, corresponding to the RCSJ model.  The element with the cross represents the Josephson element with current  related to the phase drop $\phi$ between the superconductors by $I=I_c \sin \phi$.  In the transmon configuration, the junction is shunted by a capacitor $C_s>>C$.}
    \label{fig:JJ-schematic}
\end{figure}

Our starting point is the so-called resistively and capacitively shunted junction (RCSJ) model \cite{tinkham2004introduction}, where the dynamics of the phase difference $\phi$ across the junction is determined by the equation 
\begin{equation}
    \left(\frac{\hbar}{2e}\right)^2 C \frac{d^2 \phi}{dt^2} + \left(\frac{\hbar}{2e}\right)^2 \frac{1}{R}\frac{d\phi}{dt} + \frac{\hbar}{2e} I_c \sin \phi = \frac{\hbar}{2e} I,
    \label{eq:rcsj}
\end{equation}
where $\hbar$ is the reduced Planck's constant, $e$ the electron charge, $t$ the time, $C$ the intrinsic capacitance of the junction, $R$ the shunt resistance in parallel with $C$, $I_c$ is the critical current of the junction and $I$ the applied current bias (see Fig. \ref{fig:JJ-schematic}).  $R$ here includes the quasiparticle resistance \cite{orr_phase_1986} as well as any contributions arising from the measuring setup, such as the source resistance of the driving electronics \cite{kautz_noise-affected_1990, joyez_josephson_1999, hu_low-voltage_1992} 

As has been noted, this equation is identical to that for a particle of mass $m=(\hbar/2e)^2 C$ in a tilted washboard potential $U = -E_J \cos \phi - (\hbar/2e) I \phi$, $E_J = (\hbar/2e) I_c$ being the Josephson energy, in the presence of dissipation that couples to the velocity of the particle through a parameter $\gamma = (1/RC)$
\begin{equation}
    m \Ddot{x} + m \gamma \dot{x} + U'(x) = 0
    \label{eq:classical}
\end{equation}
In analogy with the Brownian motion of a particle, the dissipation can be thought of as arising from the influence of forces random in time \cite{balakrishnan_elements_2021} which need to be included in Eqn. (\ref{eq:classical}).  A more general form of this equation including these random forces can then be written as a quantum Langevin equation \cite{ford_dissipative_1988}
\begin{equation}
    m \Ddot{x} + \int_0^\infty \mu(t-t') \dot{x}(t') dt' + U'(x) = F(t)
    \label{eq:quantumlangevin}
\end{equation}
where the autocorrelation of the random force $F(t)$ is related to the real part of the Fourier transform of the so-called memory function $\mu$ by \cite{ford_dissipative_1988}
\begin{align*}
     \frac{1}{2}&<F(t) F(t') + F(t') F(t)> \\
     &= \frac{1}{\pi} \int_0^\infty \Re[\Tilde{\mu}(\omega)] \; \hbar \omega \coth(\hbar \omega/2 k_B T) \cos\omega(t-t').
    \label{eq:correlator}   
\end{align*}

In the context of the RCSJ model, $F(t)$ would correspond to the current noise fluctuations $\delta I(t)$ introduced by the shunt resistor $R$ \cite{kautz_noise-affected_1990, joyez_josephson_1999}, but the more general formulation here could also be applied to other cases with a more complex environmental impedance, or where the response of the system depends in a more complicated way on past history \cite{hu_low-voltage_1992}, hence the moniker `memory function' for $\mu$.

The solution of Eqn. \ref{eq:quantumlangevin} is particularly illuminating in frequency space.  To simplify the analysis, we consider the case of a phase particle in the tilted washboard potential.  For small values of the current $I$, there will be local minima of the potential at $\phi = \arcsin{I/I_c}$ mod $2 \pi$, around which the potential is locally quadratic.  The potential about this minimum can be modeled as a harmonic oscillator potential $U(x) = U_0+ (1/2) m \omega_{I0}^2 x^2$, where the current-dependent harmonic oscillator frequency is given by \cite{ford_dissipative_1988} 
\begin{equation}
    \omega_{I0}^2 = \omega_0^2 \sqrt{1 - (I/I_c)^2}.
    \label{eq:current-frequency}
\end{equation}
Here $\omega_0$ is the frequency for $I=0$, and is given by $\hbar \omega_0 = \sqrt{4 e^2 E_J/C} = \sqrt{8 E_c E_J}$, where $E_c=e^2/2C$ is the single electron charging energy.

Fourier transforming Eqn. \ref{eq:quantumlangevin}, the solution can be represented as $\Tilde{x}(\omega) = \alpha(\omega) \Tilde{F}(\omega)$, where $\Tilde{x}(\omega)$ and $\Tilde{F}(\omega)$ are the Fourier transforms of $x(t)$ and $F(t)$ respectively, and the response function $\alpha(\omega)$ is given by \cite{ford_dissipative_1988}
\begin{equation}
    \alpha(\omega) = \frac{1}{-m\omega^2 + i \omega \Tilde{\mu}(\omega) + m \omega_{I0}^2}.
    \label{eq:responsefunc}
\end{equation}
As usual, the poles of $\alpha(\omega)$ give the excitation energies of the system.  For our problem, $\Tilde{\mu}(\omega)$ can be related to the environmental impedance coupled to the JJ, $\Tilde{\mu}(\omega)=(\hbar/2e)^2 Y(\omega)$, where $Y(\omega)$ is the effective admittance of the environment connected in parallel to the JJ \cite{hu_low-voltage_1992, joyez_josephson_1999}.

As an example, consider the case for $I=0$ when there is only a shunt capacitor $C_s$ connected across the JJ, as is done for a qubit in the transmon configuration (see Fig. \ref{fig:JJ-schematic}).  In this case, $Y(\omega) = i \omega C_s$.  It is easy to show that the excitation energy is then modified to $\omega_0 = \sqrt{4e^2 E_J/C} \rightarrow  \sqrt{4e^2 E_J/C_t}$, where $C_t = C_s + C$ is the parallel combination of the intrinsic capacitance $C$ and the shunt capacitance $C_s$, as expected. If we now include also ohmic dissipation in the form of a resistor $R$ connected in parallel with the junction as shown in Fig. \ref{fig:JJ-schematic}, $Y(\omega)$ is now modified to $Y(\omega) = (1/R) + i\omega C_s$ and the oscillation frequency is modified to $\sqrt{\omega_{I0}^2 - \gamma^2/4}$, where $\gamma=1/RC_t$.  The ohmic dissipation also introduces a finite lifetime of $\tau = 2/\gamma$.  Classically, this would be the time scale over which the particle stops oscillating and comes to rest at the local minimum of the tilted washboard potential.  One should note that is the behaviour of the \textit{average} position or velocity: the instantaneous position will continue to exhibit random fluctuations centered about the minimum in potential driven by noise, with corresponding fluctuations in the instantaneous velocity about zero. 

Quantum mechanically, one expects to have a series of quantized energy levels, with the ground state energy $\hbar \omega_{I0}/2$.  One can think of the dissipation due to noise as leading to a finite lifetime for the state arising from transitions to other states.  For a particle in the ground state in a tilted washboard potential, this would occur through transitions out of the local minimum to states in neighboring local minima, which might occur through thermal activation or quantum tunneling \cite{caldeira_quantum_1983, ambegaokar_quantum_1982}.  This of course is the model that has been used to describe the stochastic switching of a current-biased JJ out of the zero-voltage state at currents close to the critical current $I_c$.  In this case, once the particle tunnels out of the local minimum, it then continues down the tilted washboard potential in the so-called free-running state.  Consequently, the phase is continually evolving, and a finite voltage results.  

Here we are concerned with the low bias regime, where the current $I$ is close to zero.  There may still be a finite probability of transition to neighboring minima, but the presence of dissipation means that the particle loses its small amount of excess energy after passage to the nearest minimum, or perhaps after traversing a few such minima, and so never goes into the free-running state.  However, this process of phase diffusion will still result in a finite voltage across the junction.  For simplicity, let us consider the case when it stops at the neighboring minimum.   Each such transition would result in a phase change of $2 \pi$; if the rate of such transitions is $\Gamma$, then the average $d \phi/dt = 2 \pi \Gamma$ would result in a finite voltage $(\hbar/2e) d \phi/dt = (h/2e) \Gamma$.        

Both thermal activation and quantum tunneling can lead to transitions.  To calculate the tunneling rate, I use the WKB approximation to calculate the tunneling probability $P$ \cite{merzbacher}
\begin{equation}
    P = A e^{-2 \int_a^b \kappa(x) dx},
    \label{eq:tunnel}
\end{equation}
where
\begin{equation}
    \kappa(x) = \sqrt{\frac{2m(U(x)-E)}{\hbar^2}}.
    \label{eq:kappa}
\end{equation}
Here $A$ is a constant that is determined by the normalization of the wavefunction in the WKB approximation and is essentially unity, $E=\hbar \omega_{I0}/2$, and $a$ and $b$ are the positions in the potential where $E=U(x)$.  The probability of thermal activation over the barrier for a state at energy $E$ for $I\sim 0$ should go as $\sim e^{-(2 E_J - E)/k_B T}$ at a temperature $T$: here we consider temperatures where this term is much smaller than the tunneling term Eqn. \ref{eq:tunnel}.  The transition rate is then given by the attempt frequency multiplied by $P$.  Following Gamow \cite{Gamow1928ZurQD}, we take the attempt frequency to be $\omega_{I0}/2$, so that $\Gamma = \omega_{I0} P/2$.  For small currents, the particle can tunnel to the left and right with rates $\Gamma_l$ and $\Gamma_r$.  The resulting voltage is then $V=(h/2e)(\Gamma_r-\Gamma_l)$.

There are two relevant energy scales in the problem, the charging energy $E_c$ and the Josephson energy $E_J$ with the results depending on the ratio $E_c/E_J$.  For a fixed $E_J$, a larger $E_c$ corresponds classically to a particle with a smaller mass and thus one more susceptible to noise; quantum mechanically, since the energy of the state increases with respect to the barrier, the particle is more likely to tunnel.  We should therefore expect the voltage to increase with increasing $E_c/E_J$, as has been verified in recent experiments \cite{lu_phase_2023}.  Dissipation reduces the energy of the state, hence we expect increased dissipation should reduce the probability of tunneling, in line with previous analyses \cite{caldeira_quantum_1983}, and therefore reduce the voltage.
\begin{figure}
    \centering
    \includegraphics[width=\columnwidth]{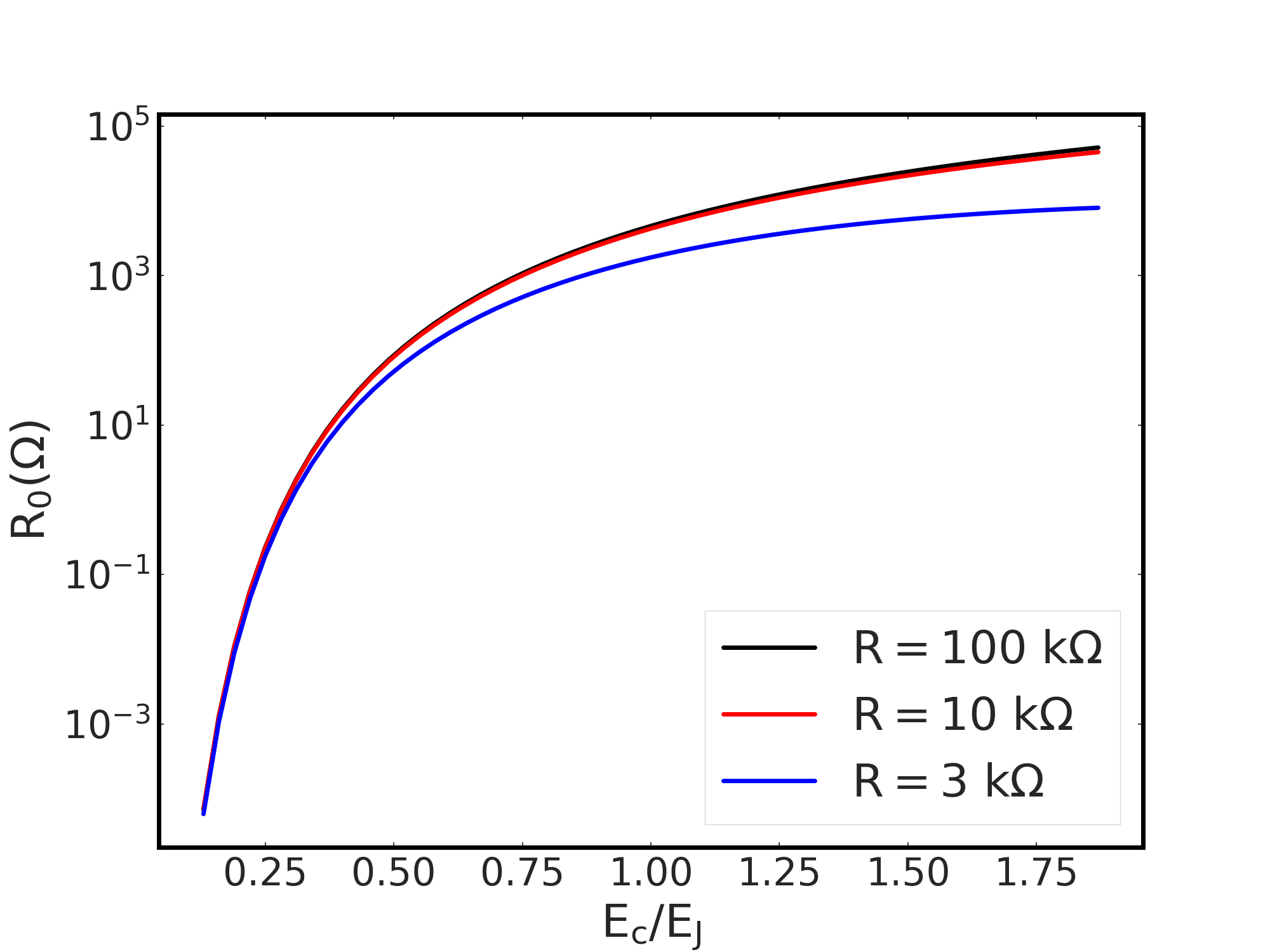}
    \caption{The calculated zero bias resistance due to quantum tunneling as a function of $E_c/E_J$ for three different values of the resistance $R$.  $I_c$ here is assumed to be 30 nA, which sets the value of $E_J$}
    \label{fig:R0_vs_Ec}
\end{figure}

At $I=0$, $V=0$, but the zero bias resistance $R_0 = dV/dI$ can be finite and increases rapidly with $E_c/E_J$.  This is shown in Fig. \ref{fig:R0_vs_Ec} for three different values of the shunt resistance $R$, assuming a value of $I_c=30$ nA, consistent with design parameters on recent transmon qubits \cite{wisne_jj_2024}.  For $E_c/E_J < 0.5$, the dependence is superexponential, with $R_0$ dropping many orders of magnitude over a narrow range of $E_c/E_J$.  For $E_c/E_J>1$ the dependence is weaker but still exponential.  Changing the shunt resistance $R$ seems to have little effect until $R$ is of the order of a few k$\Omega$ or less, and even then the major effect is at larger values of $E_c/E_J$.  A similar calculation using thermal activation instead of tunneling to generate the phase jumps gives $R_0$ values many orders of magnitude smaller than those shown in Fig, \ref{fig:R0_vs_Ec}, assuming a temperature of 25 mK, showing that phase jumps due to thermal activation can be neglected at these temperatures.  These results are similar to those obtained earlier by Hu and O'Connell \cite{hu_low-voltage_1992}.
\begin{figure}
    \centering
    \includegraphics[width=\columnwidth]{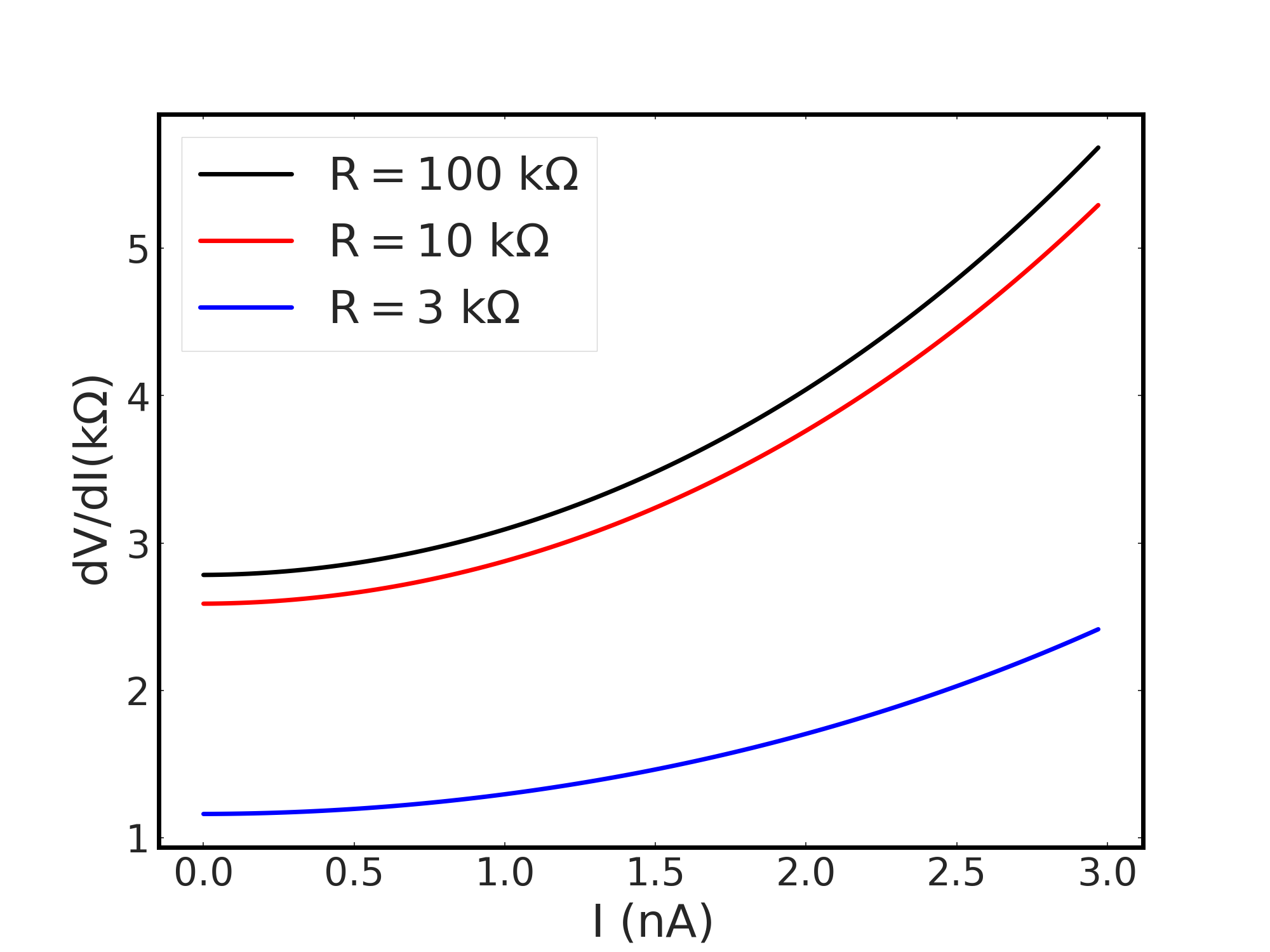}
    \caption{Differential resistance $dV/dI$ of a Josephson junction as a function of bias current $I$ for three different values of the shunt resistance $R$.  $E_c/E_J=0.9$. }
    \label{fig:dVdI_vs_I}
\end{figure}

Let us now consider the dependence of the differential resistance $dV/dI$ as a function of $I$ near the zero bias regime.  (At larger values of $I$, the JJ will transition to the free-running state, which we do not concern ourselves with here.)  Figure \ref{fig:dVdI_vs_I} shows the differential resistance as a function of bias current for three different values of the shunt resistor $R$, for a specific value of $E_c/E_J=0.9$.  There are two competing effects that modify the tunneling as the bias current is increased.  The height of the barrier decreases as $I$ is increased, which might be expected to give a rapid increase in the differential resistance as it is easier for the phase particle to transition to a neighboring minimum.  However, the energy of the ground state also decreases, which counterbalances somewhat the decrease in barrier height, leading to only a relatively gradual increase in resistance with bias.  As seen earlier above, for a fixed value of $E_c/E_J$ in this range, the overall value of the differential resistance decreases when the shunt resistance $R$ drops below a few k$\Omega$.  One should also note that the overall curvature of $dV/dI$ vs. $I$ also decreases as $R$ falls below a few k$\Omega$.  These curves are similar to those obtained on small Josephson junctions \cite{wisne_jj_2024}.  

In principle, one should be able to determine the effective capacitance and shunt resistance in the measurement by fitting measurements of $dV/dI$ vs $I$ to this model.  In reality, however, the detailed impedance environment of the Josephson junction will likely be different from the simple model we have assumed.  The power of using the quantum Langevin approach is that a variety of enviromental impedances can be modeled readily by appropriately specifying the admittance $Y(\omega)$, as has been discussed earlier.  

While we have focused here on dc electrical transport Josephson junctions, the parameters that determine their sensitivity to noise should also be applicable in the microwave regime, the usual frequency regime for superconducting qubits.  It should be possible to include various noise sources by modeling them as components of a frequency dependence admittance $Y(\omega)$.  In particular, it might be possible to model the influence of two-level-systems (TLSs), which have been identified as a major source of decoherence in superconducting qubits \cite{Kjaergaard_2020}.

This work is supported by the U.S. Department of Energy, Office of Science, National Quantum Information Science Research Centers, Superconducting Quantum Materials and Systems Center (SQMS) under Contract No. DEAC02-07CH11359.

\bibliographystyle{elsarticle-num}

\begin{thebibliography}{10}
\expandafter\ifx\csname url\endcsname\relax
  \def\url#1{\texttt{#1}}\fi
\expandafter\ifx\csname urlprefix\endcsname\relax\def\urlprefix{URL }\fi
\expandafter\ifx\csname href\endcsname\relax
  \def\href#1#2{#2} \def\path#1{#1}\fi

\bibitem{voss_macroscopic_1981}
R.~F. Voss, R.~A. Webb, Macroscopic {Quantum} {Tunneling} in 1-$\mu$m {Nb}
  {Josephson} {Junctions}, Phys. Rev. Lett. 47~(4) (1981) 265--268.
\newblock \href {https://doi.org/10.1103/PhysRevLett.47.265}
  {\path{doi:10.1103/PhysRevLett.47.265}}.

\bibitem{devoret_measurements_1985}
M.~H. Devoret, J.~M. Martinis, J.~Clarke, Measurements of {Macroscopic}
  {Quantum} {Tunneling} out of the {Zero}-{Voltage} {State} of a
  {Current}-{Biased} {Josephson} {Junction}, Phys. Rev. Lett. 55~(18) (1985)
  1908--1911.
\newblock \href {https://doi.org/10.1103/PhysRevLett.55.1908}
  {\path{doi:10.1103/PhysRevLett.55.1908}}.

\bibitem{clarke_quantum_1988}
J.~Clarke, A.~N. Cleland, M.~H. Devoret, D.~Esteve, J.~M. Martinis, Quantum
  {Mechanics} of a {Macroscopic} {Variable}: {The} {Phase} {Difference} of a
  {Josephson} {Junction}, Science 239~(4843) (1988) 992--997.
\newblock \href {https://doi.org/10.1126/science.239.4843.992}
  {\path{doi:10.1126/science.239.4843.992}}.

\bibitem{ambegaokar_quantum_1982}
V.~Ambegaokar, U.~Eckern, G.~Schön, Quantum {Dynamics} of {Tunneling} between
  {Superconductors}, Phys. Rev. Lett. 48~(25) (1982) 1745--1748.
\newblock \href {https://doi.org/10.1103/PhysRevLett.48.1745}
  {\path{doi:10.1103/PhysRevLett.48.1745}}.

\bibitem{fisher_quantum_1985}
M.~P.~A. Fisher, W.~Zwerger, Quantum {Brownian} motion in a periodic potential,
  Phys. Rev. B 32~(10) (1985) 6190--6206.
\newblock \href {https://doi.org/10.1103/PhysRevB.32.6190}
  {\path{doi:10.1103/PhysRevB.32.6190}}.

\bibitem{clarke_macroscopic_1986}
J.~Clarke, G.~Schön, Macroscopic {Quantum} {Phenomena} in {Josephson}
  {Elements}, Europhys. News 17~(7-8) (1986) 94--96.
\newblock \href {https://doi.org/10.1051/epn/19861707094}
  {\path{doi:10.1051/epn/19861707094}}.

\bibitem{blackburn_survey_2016}
J.~A. Blackburn, M.~Cirillo, N.~Grønbech-Jensen, A survey of classical and
  quantum interpretations of experiments on {Josephson} junctions at very low
  temperatures, Physics Reports 611 (2016) 1--33.
\newblock \href {https://doi.org/10.1016/j.physrep.2015.10.010}
  {\path{doi:10.1016/j.physrep.2015.10.010}}.

\bibitem{caldeira_quantum_1983}
A.~O. Caldeira, A.~J. Leggett, Quantum tunnelling in a dissipative system,
  Annals of Physics 149~(2) (1983) 374--456.
\newblock \href {https://doi.org/10.1016/0003-4916(83)90202-6}
  {\path{doi:10.1016/0003-4916(83)90202-6}}.

\bibitem{iansiti_charging_1987}
M.~Iansiti, A.~T. Johnson, W.~F. Smith, H.~Rogalla, C.~J. Lobb, M.~Tinkham,
  Charging energy and phase delocalization in single very small {Josephson}
  tunnel junctions, Phys. Rev. Lett. 59~(4) (1987) 489--492.
\newblock \href {https://doi.org/10.1103/PhysRevLett.59.489}
  {\path{doi:10.1103/PhysRevLett.59.489}}.

\bibitem{iansiti_crossover_1988}
M.~Iansiti, A.~T. Johnson, C.~J. Lobb, M.~Tinkham, Crossover from {Josephson}
  {Tunneling} to the {Coulomb} {Blockade} in {Small} {Tunnel} {Junctions},
  Phys. Rev. Lett. 60~(23) (1988) 2414--2417.
\newblock \href {https://doi.org/10.1103/PhysRevLett.60.2414}
  {\path{doi:10.1103/PhysRevLett.60.2414}}.

\bibitem{kautz_noise-affected_1990}
R.~L. Kautz, J.~M. Martinis, Noise-affected \textit{{I}} - \textit{{V}} curves
  in small hysteretic {Josephson} junctions, Phys. Rev. B 42~(16) (1990)
  9903--9937.
\newblock \href {https://doi.org/10.1103/PhysRevB.42.9903}
  {\path{doi:10.1103/PhysRevB.42.9903}}.

\bibitem{martinis_classical_1989}
J.~M. Martinis, R.~L. Kautz, Classical phase diffusion in small hysteretic
  {Josephson} junctions, Phys. Rev. Lett. 63~(14) (1989) 1507--1510.
\newblock \href {https://doi.org/10.1103/PhysRevLett.63.1507}
  {\path{doi:10.1103/PhysRevLett.63.1507}}.

\bibitem{joyez_josephson_1999}
P.~Joyez, D.~Vion, M.~Götz, M.~H. Devoret, D.~Esteve, The {Josephson} {Effect}
  in {Nanoscale} {Tunnel} {Junctions}, Journal of Superconductivity 12~(6)
  (1999) 757--766.
\newblock \href {https://doi.org/10.1023/A:1007733009637}
  {\path{doi:10.1023/A:1007733009637}}.

\bibitem{ivanchenko_josephson_1969}
Y.~M. Ivanchenko, A.~Zil'Berman, {THE} {JOSEPHSON} {EFFECT} {IN} {SMALL}
  {TUNNEL} {CONTACTS}, Sov. Phys. JETP. (1969) 1272.

\bibitem{biswas_effect_1970}
A.~C. Biswas, S.~S. Jha, Effect of {Thermal} {Noise} on the dc {Josephson}
  {Effect}, Phys. Rev. B 2~(7) (1970) 2543--2547.
\newblock \href {https://doi.org/10.1103/PhysRevB.2.2543}
  {\path{doi:10.1103/PhysRevB.2.2543}}.

\bibitem{hu_low-voltage_1992}
G.~Y. Hu, R.~F. O'Connell, Low-voltage resistance in small {Josephson}
  junctions, J. Phys.: Condens. Matter 4~(48) (1992) 9635--9642.
\newblock \href {https://doi.org/10.1088/0953-8984/4/48/017}
  {\path{doi:10.1088/0953-8984/4/48/017}}.

\bibitem{buttiker_thermal_1983}
M.~Büttiker, E.~P. Harris, R.~Landauer, Thermal activation in extremely
  underdamped {Josephson}-junction circuits, Phys. Rev. B 28~(3) (1983)
  1268--1275.
\newblock \href {https://doi.org/10.1103/PhysRevB.28.1268}
  {\path{doi:10.1103/PhysRevB.28.1268}}.

\bibitem{merzbacher}
E.~Merzbacher, Quantum Mechanics, 2nd Edition, John Wiley \& Sons, 1961.

\bibitem{tinkham2004introduction}
M.~Tinkham, Introduction to Superconductivity, 2nd Edition, Dover Publications,
  2004.

\bibitem{orr_phase_1986}
B.~G. Orr, J.~R. Clem, H.~M. Jaeger, A.~M. Goldman, Phase fluctuations in
  {Josephson} junctions, Phys. Rev. B 34~(5) (1986) 3491--3494.
\newblock \href {https://doi.org/10.1103/PhysRevB.34.3491}
  {\path{doi:10.1103/PhysRevB.34.3491}}.

\bibitem{balakrishnan_elements_2021}
{Elements} {of} {Nonequilibrium} {Statistical} {Mechanics}, Cham.

\bibitem{ford_dissipative_1988}
G.~Ford, J.~Lewis, R.~O'Connell, Dissipative quantum tunneling: quantum
  {Langevin} equation approach, Physics Letters A 128~(1-2) (1988) 29--34.
\newblock \href {https://doi.org/10.1016/0375-9601(88)91037-7}
  {\path{doi:10.1016/0375-9601(88)91037-7}}.

\bibitem{Gamow1928ZurQD}
G.~Gamow, Zur quantentheorie des atomkernes, Zeitschrift f{\"u}r Physik 51
  (1928) 204--212.

\bibitem{lu_phase_2023}
W.-S. Lu, K.~Kalashnikov, P.~Kamenov, T.~J. DiNapoli, M.~E. Gershenson, Phase
  {Diffusion} in {Low}-{EJ} {Josephson} {Junctions} at {Milli}-{Kelvin}
  {Temperatures}, Electronics 12~(2) (2023) 416.
\newblock \href {https://doi.org/10.3390/electronics12020416}
  {\path{doi:10.3390/electronics12020416}}.

\bibitem{wisne_jj_2024}
M.~Wisne, in preparation (2024).

\bibitem{Kjaergaard_2020}
M.~Kjaergaard, M.~E. Schwartz, J.~Braum\"{u}ller, P.~Krantz, J.~I.-J. Wang,
  S.~Gustavsson, W.~D. Oliver, Superconducting qubits: Current state of play,
  Annual Review of Condensed Matter Physics 11~(1) (2020) 369--395.
\newblock \href
  {http://arxiv.org/abs/https://doi.org/10.1146/annurev-conmatphys-031119-050605}
  {\path{arXiv:https://doi.org/10.1146/annurev-conmatphys-031119-050605}},
  \href {https://doi.org/10.1146/annurev-conmatphys-031119-050605}
  {\path{doi:10.1146/annurev-conmatphys-031119-050605}}.

\end{thebibliography}
 \newcommand{\noop}[1]{}

\end{document}